\documentclass[a4paper,11pt]{article}
\usepackage{times}
\usepackage{geometry}             
\usepackage{graphicx}
\usepackage{amssymb}
\usepackage{epstopdf}
\usepackage{natbib}
\long\def\symbolfootnote[#1]#2{\begingroup%
\def\thefootnote{\fnsymbol{footnote}}\footnote[#1]{#2}\endgroup}

\def\gs{\mathrel{\raise0.35ex\hbox{$\scriptstyle >$}\kern-0.6em
\lower0.40ex\hbox{{$\scriptstyle \sim$}}}}
\def\ls{\mathrel{\raise0.35ex\hbox{$\scriptstyle <$}\kern-0.6em
\lower0.40ex\hbox{{$\scriptstyle \sim$}}}}

\begin{document}

\pagenumbering{arabic}

\noindent{\small{ \it \bf ``UKIRT at 30 -- A British Success Story'', Royal Observatory of Edinburgh, 14-16 September, 2009}}
\bigskip
\bigskip

\noindent{\LARGE \bf  The HiZELS/UKIRT large survey for bright Ly$\alpha$ emitters at $z\sim9$\symbolfootnote[1]{In this work, an H$_0=70$\,km\,s$^{-1}$\,Mpc$^{-1}$, $\Omega_M=0.3$ and $\Omega_{\Lambda}=0.7$ cosmology is used and magnitudes are given in the Vega system. For full details, please refer to Sobral et al. (2009b).} } 

\bigskip
\bigskip
{\addtolength{\baselineskip}{2pt}
\noindent{\Large{\it David Sobral$^1$, Philip Best$^1$, Jim Geach$^2$, Ian Smail$^2$, Jaron Kurk$^3$, Michele Cirasuolo$^4$,}}

\noindent{\Large{\it Mark Casali$^5$, Rob Ivison$^4$, Kristen Coppin$^2$ \& Gavin Dalton$^6$}}

\smallskip

\noindent{\footnotesize{\it 1- IfA, Edinburgh University, UK; $2$- ICC, Durham University, UK; $3$- MPE, Garching, Germany; $ 4$- ESO, Garching, Germany;}}

\noindent{\footnotesize{\it $5$- Oxford University, UK}}
\bigskip

%
%
%
\centerline{\Large{\bf Abstract}}
\bigskip

\noindent We present the largest area survey to date (1.4 deg$^2$) for Ly$\alpha$ emitters (LAEs) at $z\sim9$, as part of the Hi-$z$ Emission Line Survey (HiZELS). The survey, which primarily targets H$\alpha$ emitters at $z<3$, uses the Wide Field CAMera on the United Kingdom Infrared Telescope and a custom narrow-band filter in the $J$ band to reach a Ly$\alpha$ luminosity limit of $\sim10^{43.8}$\,erg\,s$^{-1}$ over a co-moving volume of $1.12\times10^6$\,Mpc$^3$ at $z=8.96\pm0.06$. Two candidates were found out of 1517 line emitters, but those were rejected as LAEs after follow-up observations. This improves the limit on the space density of bright Ly$\alpha$ emitters by 3 orders of magnitude and is consistent with suppression of the bright end of the Ly$\alpha$ luminosity function beyond $z\sim6$. Combined with upper limits from smaller but deeper surveys, this rules out some of the most extreme models for high-redshift Ly$\alpha$ emitters. The potential contamination of narrow-band Ly$\alpha$ surveys at $z>7$ by Galactic brown dwarf stars is also examined, leading to the conclusion that such contamination may well be significant for searches at $7.7<z<8.0$, $9.1<z<9.5$ and $11.7 < z < 12.2$.

\section{Introduction}

Understanding how and when the first stars and galaxies formed is one of the most fundamental problems in astronomy. Furthermore, whilst many sophisticated models of early galaxy formation and evolution have been constructed, it is clear that observations of the most distant galaxies are mandatory to really test, refine, or refute such models. Indeed, considerable manpower and telescope time have been invested in such observations, with the detection of a Gamma Ray Burst (GRB) at $z\approx 8.2$ \citep[][]{Tanvir} being one of the most recent highlights of this extraordinary endeavor. However, despite the recent success in using GRBs to find the most distant sources, the current samples of high redshift galaxies have been mostly assembled using two methods: the broad-band drop-out technique and narrow-band imaging surveys.

The widely-used drop-out technique \citep[pioneered at $z\sim3$ by][]{steidel96} requires very deep broad-band imaging, and can identify $z>7$ galaxies as z-band drop-outs \citep[e.g.][]{bouwens08,richard08}. Furthermore, the use of this technique, combined with the recent installation of the Wide Field Camera 3 (WFC3) on the Hubble Space Telescope (HST), has led to the identification of roughly 20 $z\approx7-8$ candidates \citep[e.g.][]{2009arXiv0909.1803B,Oesch,McLure09}. While this is an efficient method for identifying candidates, it still requires detailed spectroscopic follow-up to confirm them, especially to rule out contributions from other populations with large z$-J$ breaks, such as dusty or evolved $z\sim2$ galaxies and ultra-cool galactic stars \citep[e.g.][]{mclure06}. Confirming the candidates is actually quite a significant challenge, since the typical $z>7$ candidates found so far are just too faint for spectroscopic follow-up. 

The narrow-band imaging technique has the advantage of probing very large volumes looking for Ly$\alpha$ in emission. Whilst it can only detect sources with strong emission lines, and still depends on the Lyman-break technique to isolate very high-redshift emitters, it can yield suitable targets for follow-up spectroscopy with the current instrumentation. Narrow-band Ly$\alpha$ searches at $3<z<7$ have been extremely successful in detecting and confirming emitters \citep[e.g.][]{hu99} and, so far, this technique has resulted in the spectroscopic confirmation of the highest redshift galaxy \citep[$z=6.96$:][]{Iye}. Even more recently, \cite{Hibon} identified 7 candidate Ly$\alpha$ emitters at $z=7.7$. There have been attempts to detect Ly$\alpha$ emitters at $z\sim9$ \citep[e.g.][]{willis05,cuby07,willis08}, but all such studies have been unsuccessful to date, having surveyed very small areas (a few tens of square arcmins at most). 

With the advent of wide-field near-IR detectors it is now possible to increase the sky area studied by over 2 to 3 orders of magnitude and reach the regime where one can realistically expect to detect $z \sim 9$ objects. This is a key aim of, for example, the narrow-band component of the UltraVISTA Survey \citep[c.f.][]{Nilsson07}. It is also an aim of HiZELS, the Hi-Z Emission Line Survey \citep[c.f.][]{Geach,Sobral,Sobral2,Garn}, that we are carrying out using the WFCAM instrument on the 3.8-m UK Infrared Telescope (UKIRT). HiZELS is using a set of existing and custom-made narrow-band filters in the $J$, $H$ and $K$ bands to detect emission lines from galaxies at different redshifts over $\sim$ 10 square degrees of extragalactic sky. In particular, the narrow-band $J$ filter (hereafter NB$_{\rm J}$) is sensitive to Ly$\alpha$ at $z=8.96$.

\section{Data, selection and candidates}

Deep narrow-band $J$ (NB$_{\rm J}\approx21.6$, 3$\sigma$, ${\rm F_{lim}}=7.6\times10^{-17}$\,erg\,s$^{-1}$\,cm$^{-2}$) imaging was obtained across 1.4 deg$^2$ in the UKIRT Infrared Deep Sky Survey Ultra Deep Survey \citep[UKIDSS UDS;][]{2007MNRAS.379.1599L} and the Cosmological Evolution Survey \citep[COSMOS;][]{2007ApJS..172....1S} fields, both of which have a remarkable set of deep multi-wavelength data available -- this resulted in the selection of 1517 potential line emitters. The NB$_{\rm J}$ filter ($\lambda=1.211\pm0.015\mu$m) is sensitive to Ly$\alpha$ emission at $z=8.96\pm0.06$, probing a co-moving volume of 1.12$\times10^6$ Mpc$^3$ -- by far the largest probed by a narrow-band survey at these wavelengths. Details regarding the observations, data reduction and the general selection of NB$_{\rm J}$ emitters can be found in \cite{Sobral}.

For a source to be considered a candidate $z\approx9$ Ly$\alpha$ emitter it is required to: i) be selected as a narrow-band emitter in \cite{Sobral}; ii) have at least one other detection $>3\sigma$ in the near-infrared; iii) be visually believable in NB$_{J}$ and the other band(s), avoiding noisy areas; and iv) be undetected ($<3\sigma$ and direct visual analysis) in the available visible band imaging ($B$,$V$,$r$,$i$,z) -- \sc{subaru} \rm and ACS/\sc{hst}.

\rm No candidates were found in the UKIDSS UDS field, with all emitters that passed tests i) to iii) being clearly detected in z-band imaging. In COSMOS, however, 2 candidates were found that satisfied all criteria. The brightest source was followed-up spectroscopically using the CGS4 instrument on UKIRT in January 2009 -- these data failed to confirm an emission line. Both candidates were then re-observed using WFCAM (further $J$ imaging in February 2009), resulting in the non-detection of both candidates. Further investigation shows that the sources are likely to be artifacts caused by an unfortunate coincidence of a set of slightly hot pixels (not sufficient to be flagged as bad pixels) which, combined with the ditter pattern, produced a few $\sigma$ excess at one location on the combined image.

\section{Ly$\alpha$ luminosity function at $\bf z\sim9$}

A non-detection of (${\rm L}>7.6\times10^{43}$\,erg\,s$^{-1}$) Ly$\alpha$ emitters at $z\sim9$, in a co-moving volume of 1.12$\times10^6$ Mpc$^3$ allows the tightest constraint on the bright end of the $z\sim9$ Ly$\alpha$ luminosity function, as previous surveys \citep{willis05,cuby07,willis08} have only covered very small areas (a factor $\sim$1000 smaller). However, since those surveys have gone significantly deeper (up to a factor of $\sim$100), combining all the results from the literature can constrain the Ly$\alpha$ luminosity function across a wide range of luminosities (10$^{42}<L<10^{45}$\,erg\,s$^{-1}$). The left panel of Figure \ref{fig1} presents such constraints, indicating the inverse of the volume selection function for each survey. These are compared to the measured Ly$\alpha$ luminosity functions from $z\sim3$ to $z\sim7$, revealing that there is little evolution in the bright end of the luminosity function between $z\sim3$ and $z\sim5.7$. Nevertheless, those bright emitters seem to become much rarer at $z=6.5$ \citep{Kashikawa06}, indicating that $L^*$ is not increasing from $z\sim6$ onwards. The results presented here are also consistent with no (or negative) evolution in $L^*$ ($\Delta$log(L$^*$)$<$0.5) from $z=5.7$ to $z\sim9$.

\subsection{Comparison with models and future surveys}
\label{compl}

Many authors have made predictions regarding the Ly$\alpha$ luminosity function at $z\sim9$, either by extrapolating the luminosity function of these emitters from lower redshift, or by using numerical or semi-analytical models. The semi-analytical models discussed here are obtained from {\sc galform} \citep{Baugh05} -- these are based on $\Lambda$CDM, having been successful in reproducing a wide range of galaxy properties at different redshifts, including Ly$\alpha$ emitters up to $z\sim6$ \citep[c.f.][]{Baugh05,LeDelliou06,Orsi}. The observational approach, as in \cite{Nilsson07}, extrapolates the Schechter function parameters based on those obtained in the $3.1<z<6.5$ redshift range. In practice, this results in little $L^*$ evolution but a significant negative $\phi^*$ evolution. Finally, the phenomenological approach in \cite{Thommes05} assumes that Ly$\alpha$ emitters at high redshift are spheroids seen during their formation phase. Each galaxy is assumed to be visible as a Ly$\alpha$ emitter during a starburst phase of fixed duration that occurs at a specific redshift, drawn from a broad distribution \citep[c.f.][]{Thommes05}.

\begin{figure*}
\begin{minipage}[b]{0.48\linewidth} 
\centering
\includegraphics[width=8.2cm]{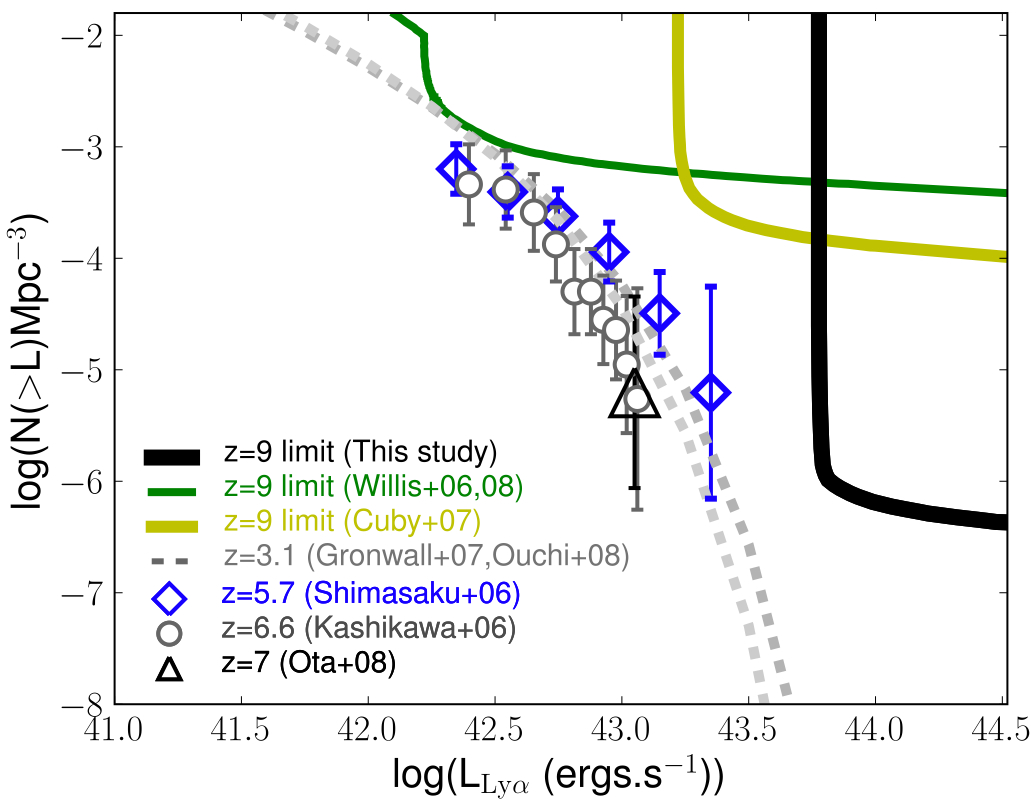}
\end{minipage}
\hspace{0.1cm} 
\begin{minipage}[b]{0.48\linewidth}
\centering
\includegraphics[width=8.2cm]{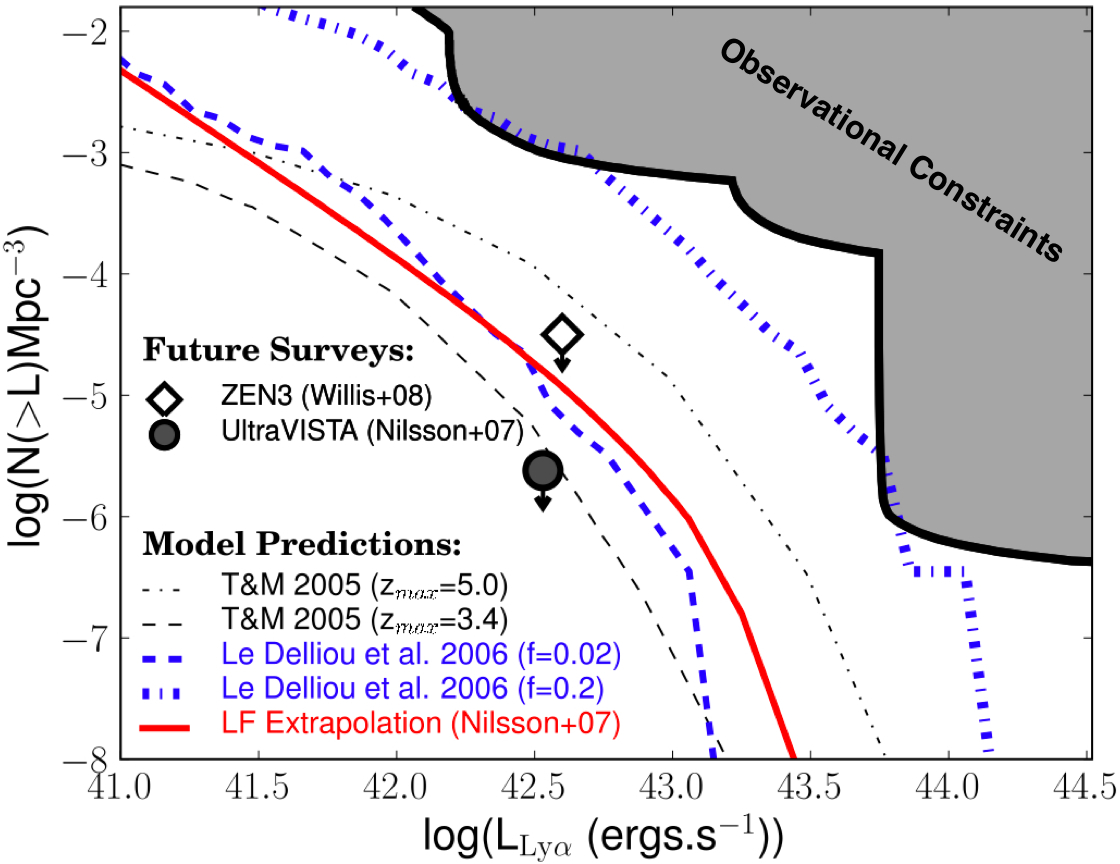}
\end{minipage}
\caption[LF1]{\footnotesize $Left$: Comparison between the measured Ly$\alpha$ luminosity function at $z\sim3$ \citep[dotted lines;][]{Gronwall07,Ouchi08} with data from $z\sim6-7$ \citep{Kashikawa06,Shimasaku06,Ota08}. No evidence of significant evolution is found, especially when accounting for cosmic variance. Limits for the $z\sim9$ LF from \cite{willis05}, \cite{cuby07} and \cite{willis08} are also presented, together with the one presented in this contribution. $Right$: The observational limits on the $z\sim9$ Ly$\alpha$ luminosity function compared to different model predictions and proposed future surveys, showing that the most recent versions of these models are completely consistent with the observations. \label{fig1} \label{fig2}}
\end{figure*}

The right panel of Figure \ref{fig2} presents predictions from {\sc galform} \citep{LeDelliou06}, the observational luminosity function extrapolation from \cite{Nilsson07} and updated phenomenological predictions \citep{Thommes05} assuming peak redshifts of $z_{max}=3.4$ and $z_{max}=5.0$. While most predictions are consistent with the current limits, {\sc galform} models with high escape fractions are marginally rejected both at faint and bright levels. Earlier phenomenological models \citep[e.g. the $z_{max}=10$ model of][not shown in Figure \ref{fig2}]{Thommes05} are also clearly rejected by our results.

\section{High-redshift Ly$\alpha$ searches and cool galactic stars}

It has become widely realised in recent years that broad-band searches for $z>6$ galaxies using the Lyman-break technique may suffer from significant contamination by cool Galactic L, T, and possibly Y-dwarf stars \citep[e.g.][]{2006MNRAS.372..357M}. These low-mass brown dwarfs display extremely red $z-J$ colours reaching as high as $z-J \approx 4$ \citep[e.g.][]{2008MNRAS.391..320B}, coupled with relatively flat $J-K$ colours. Such colours can mimic very closely those expected of a $z>6$ star forming galaxy with a strong Lyman-break.

Since the near-infrared continuum spectra of low mass brown dwarfs show considerable structure due to broad molecular absorption features \citep[especially methane and ammonia; e.g.][]{2007ApJ...667..537L}, as shown in the top panel of Figure \ref{fig3}, they can easily produce a positive broad-band minus narrow-band (BB-NB) colour (see lower panel of Figure \ref{fig3}) if the narrow-band filter is located within one of the spectral peaks (this is much less of an issue for surveys which difference two closely-located narrow-band filters). Ly$\alpha$ narrow-band surveys in the redshift ranges $7.7<z<8.0$, $9.1<z<9.5$ and $11.7 < z < 12.2$ are therefore prone to contamination by cool Galactic stars -- this includes the $z=7.7$ and $z=9.4$ atmospheric windows for narrow-band searches of Ly$\alpha$ emitters. Narrow-band surveys at redshifts $z<7.7$, or between $8.0<z<9.1$ -- which include both HiZELS ($z=8.96$) and the narrow-band component of the UltraVISTA Survey ($z=8.8$; e.g. Nilsson et al 2007) -- will not only be free of such contamination, but can potentially select very cool T-dwarf stars via a narrow-band {\it deficit} (due to the strong methane absorption feature at these wavelengths). Indeed, motivated by such finding, a T-dwarf search was conducted among narrow-band {\it deficit} sources from Sobral et al. (2009a). The results show that all those sources are galaxies with $z_{\rm photo}\sim1.4-1.5$, probably placing the H$\beta$ and [O{\sc iii}] emission lines just outside the narrow-band coverage, but contributing significantly to the measured $J$ flux, which results in the observed narrow-band deficits. No T-dwarf candidate was found.

%
%
\begin{figure}
\centering
\includegraphics[width=9cm]{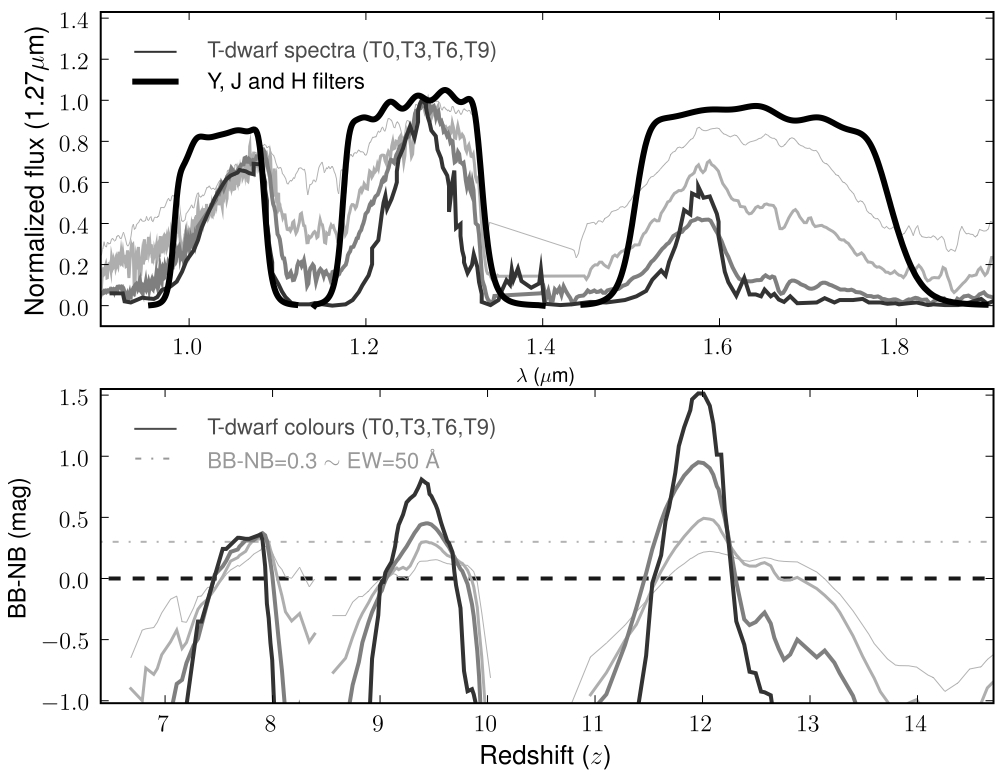}
\caption[Tdwarfs]{\footnotesize {\it Top panel:} the near-infrared spectra of T0, T3, T6 and T9 dwarf stars \citep[T0 -- lighter, T9 -- darker, from][]{2008MNRAS.391..320B} compared to near-IR broad band filter profiles. {\it Lower panel:} the consequences for measured broad-band minus narrow-band (BB-NB) colours, clearly demonstrating the redshifts/wavelengths at which searches for Ly$\alpha$ emitters can be significantly contaminated by these very cool stars. For $7.7<z<8.0$ and $9.1<z<9.5$ searches, these stars can easily mimic Ly$\alpha$ emitters, with strong $Y$-z or $J$-z breaks and significant positive BB-NB colours. Searches at higher redshift $11.6 < z < 12.2$ in the $H$ band can detect T9s with BB-NB$\sim$1.5, although the lack of strong H-J or H-Y breaks will make it easier to distinguish T-dwarfs from Ly$\alpha$ emitters. \label{fig3}}
\end{figure}

\section{Summary}

\begin{itemize}

\item Deep narrow-band imaging in the $J$ band ($\lambda=1.211\pm0.015\mu$m) has been used to search for bright Ly$\alpha$ emitters at $z=8.96$ over an area of 1.4 deg$^2$. No Ly$\alpha$ emitter was found brighter than ${\rm L}\approx7.6\times10^{43}$\,erg\,s$^{-1}$.
\item The Ly$\alpha$ luminosity function constraints at $z\sim9$ have been significantly improved for $L>10^{43.8}$\,erg\,s$^{-1}$ and combined with constraints from deeper but much smaller previous surveys. The results rule out significant positive evolution of the Ly$\alpha$ LF beyond $z\sim6$; they are in line with recent semi-analytic \& phenomenological model predictions, rejecting some extreme models.
\item It has been shown that for narrow-band searches, T-dwarfs can mimic Ly$\alpha$ emitters at $7.7<z<8.0$, $9.1<z<9.5$ and $11.7 < z < 12.2$; they will not contaminate the future UltraVISTA narrow-band survey (and can even be identified via a narrow-band {\it deficit}), but they may contaminate narrow-band Ly$\alpha$ searches within the $z=7.7$ and $z=9.4$ atmospheric windows.
\end{itemize}

These results show that bright ${\rm L}>10^{43.8}$\,erg\,s$^{-1}$ Ly$\alpha$ emitters are extremelly rare. Although the area coverage is absolutely important, a depth+area combination is likely to be the best approach for gathering the first sample of these very high-redshift galaxies. In fact, that is the strategy of the narrow-band component of the UltraVISTA survey \cite[c.f.][]{Nilsson07}, using the VISTA telescope, which will map 0.9 deg$^2$ of the COSMOS field to a planned 5$\sigma$ luminosity limit of $L=10^{42.53}$\,erg\,s$^{-1}$ and a surveyed volume of 5.41$\times10^5$ Mpc$^3$ (see right panel of Figure \ref{fig2}) at $z=8.8$. This combination lies below all current predictions for the $z\sim9$ Ly$\alpha$ LF and the survey is expected to detect 2-20 Ly$\alpha$ emitters at $z=8.8\pm0.1$. Furthermore, the continuation of HiZELS on UKIRT and the extension of the narrow-band $J$ survey to a wider area offers a complementary approach which might be able to detect one of the brightest Ly$\alpha$ emitters at $z\sim9$, perfectly suited for spectroscopic follow-up and potentially enabling the detailed studies which simply won't be possible for much fainter emitters, even if they are detected.

\bigskip

\small\bibliographystyle{mn2e.bst}
\small\bibliography{bibliography.bib}

\begin{thebibliography}{}

\bibitem[\protect\citeauthoryear{{Baugh}, {Lacey}, {Frenk}, {Granato}, {Silva}
  \& {et al.}}{{Baugh} et~al.}{2005}]{Baugh05}
{Baugh} C.~M.,  {Lacey} C.~G.,  {Frenk} C.~S.,  {Granato} G.~L.,  {Silva} L.,
   {et al.} 2005, MNRAS, 356, 1191

\bibitem[\protect\citeauthoryear{{Bouwens}, {Illingworth}, {Franx} \&
  {Ford}}{{Bouwens} et~al.}{2008}]{bouwens08}
{Bouwens} R.~J.,  {Illingworth} G.~D.,  {Franx} M.,    {Ford} H.,  2008, ApJ,
  686, 230

\bibitem[\protect\citeauthoryear{{Bouwens et al.}}{{Bouwens et
  al.}}{2009}]{2009arXiv0909.1803B}
{Bouwens et al.} 2009, arXiv:0909.1803

\bibitem[\protect\citeauthoryear{{Burningham et al.}}{{Burningham et
  al.}}{2008}]{2008MNRAS.391..320B}
{Burningham et al.} 2008, MNRAS, 391, 320

\bibitem[\protect\citeauthoryear{{Cuby}, {Hibon}, {Lidman}, {Le F{\`e}vre} \&
  {et al.}}{{Cuby} et~al.}{2007}]{cuby07}
{Cuby} J.-G.,  {Hibon} P.,  {Lidman} C.,  {Le F{\`e}vre} O.,    {et al.} 2007,
  AAP, 461, 911

\bibitem[\protect\citeauthoryear{{Garn}, {Sobral}, {Best}, {Geach}, {Smail} \&
  {et al.}}{{Garn} et~al.}{2009}]{Garn}
{Garn} T.,  {Sobral} D.,  {Best} P.~N.,  {Geach} J.~E.,  {Smail} I.,    {et
  al.} 2009, arXiv:0911.2511

\bibitem[\protect\citeauthoryear{{Geach}, {Smail}, {Best}, {Kurk}, {Casali},
  {Ivison} \& {Coppin}}{{Geach} et~al.}{2008}]{Geach}
{Geach} J.~E.,  {Smail} I.,  {Best} P.~N.,  {Kurk} J.,  {Casali} M.,  {Ivison}
  R.~J.,    {Coppin} K.,  2008, MNRAS, 388, 1473

\bibitem[\protect\citeauthoryear{{Gronwall et al.}}{{Gronwall et
  al.}}{2007}]{Gronwall07}
{Gronwall et al.} 2007, ApJ, 667, 79

\bibitem[\protect\citeauthoryear{{Hibon}, {Cuby}, {Willis}, {Cl{\'e}ment} \&
  {et al.}}{{Hibon} et~al.}{2009}]{Hibon}
{Hibon} P.,  {Cuby} J.,  {Willis} J.,  {Cl{\'e}ment} B.,    {et al.} 2009,
  arXiv:0907.3354

\bibitem[\protect\citeauthoryear{{Hu}, {Cowie} \& {McMahon}}{{Hu}
  et~al.}{1998}]{hu99}
{Hu} E.~M.,  {Cowie} L.~L.,    {McMahon} R.~G.,  1998, ApJL, 502, L99+

\bibitem[\protect\citeauthoryear{{Iye et al.}}{{Iye et al.}}{2006}]{Iye}
{Iye et al.} 2006, Nature, 443, 186

\bibitem[\protect\citeauthoryear{{Kashikawa et al.}}{{Kashikawa et
  al.}}{2006}]{Kashikawa06}
{Kashikawa et al.} 2006, ApJ, 648, 7

\bibitem[\protect\citeauthoryear{{Lawrence et al.}}{{Lawrence et
  al.}}{2007}]{2007MNRAS.379.1599L}
{Lawrence et al.} 2007, MNRAS, 379, 1599

\bibitem[\protect\citeauthoryear{{Le Delliou}, {Lacey}, {Baugh} \&
  {Morris}}{{Le Delliou} et~al.}{2006}]{LeDelliou06}
{Le Delliou} M.,  {Lacey} C.~G.,  {Baugh} C.~M.,    {Morris} S.~L.,  2006,
  MNRAS, 365, 712

\bibitem[\protect\citeauthoryear{{Leggett}, {Marley}, {Freedman}, {Saumon} \&
  {et al.}.}{{Leggett} et~al.}{2007}]{2007ApJ...667..537L}
{Leggett} S.~K.,  {Marley} M.~S.,  {Freedman} R.,  {Saumon} D.,    {et al.}.
  2007, ApJ, 667, 537

\bibitem[\protect\citeauthoryear{{McLure}, {Dunlop}, {Cirasuolo} \& {et
  al.}}{{McLure} et~al.}{2009}]{McLure09}
{McLure} R.~J.,  {Dunlop} J.~S.,  {Cirasuolo} M.,    {et al.} 2009,
  arXiv:0909.2437

\bibitem[\protect\citeauthoryear{{McLure}, {Jarvis}, {Targett}, {Dunlop} \&
  {Best}}{{McLure} et~al.}{2006}]{mclure06}
{McLure} R.~J.,  {Jarvis} M.~J.,  {Targett} T.~A.,  {Dunlop} J.~S.,    {Best}
  P.~N.,  2006, MNRAS, 368, 1395

\bibitem[\protect\citeauthoryear{{McLure et al.}}{{McLure et
  al.}}{2006}]{2006MNRAS.372..357M}
{McLure et al.} 2006, MNRAS, 372, 357

\bibitem[\protect\citeauthoryear{{Nilsson}, {Orsi}, {Lacey}, {Baugh} \&
  {Thommes}}{{Nilsson} et~al.}{2007}]{Nilsson07}
{Nilsson} K.~K.,  {Orsi} A.,  {Lacey} C.~G.,  {Baugh} C.~M.,    {Thommes} E.,
  2007, A\&A, 474, 385

\bibitem[\protect\citeauthoryear{{Oesch et al.}}{{Oesch et al.}}{2009}]{Oesch}
{Oesch et al.} 2009, arXiv:0909.1806

\bibitem[\protect\citeauthoryear{{Orsi}, {Lacey}, {Baugh} \& {Infante}}{{Orsi}
  et~al.}{2008}]{Orsi}
{Orsi} A.,  {Lacey} C.~G.,  {Baugh} C.~M.,    {Infante} L.,  2008, MNRAS, 391,
  1589

\bibitem[\protect\citeauthoryear{{Ota et al.}}{{Ota et al.}}{2008}]{Ota08}
{Ota et al.} 2008, ApJ, 677, 12

\bibitem[\protect\citeauthoryear{{Ouchi et al.}}{{Ouchi et
  al.}}{2008}]{Ouchi08}
{Ouchi et al.} 2008, ApJS, 176, 301

\bibitem[\protect\citeauthoryear{{Richard}, {Stark}, {Ellis} \& {et
  al.}}{{Richard} et~al.}{2008}]{richard08}
{Richard} J.,  {Stark} D.~P.,  {Ellis} R.~S.,    {et al.} 2008, ApJ, 685, 705

\bibitem[\protect\citeauthoryear{{Scoville et al.}}{{Scoville et
  al.}}{2007}]{2007ApJS..172....1S}
{Scoville et al.} 2007, APJS, 172, 1

\bibitem[\protect\citeauthoryear{{Shimasaku et al.}}{{Shimasaku et
  al.}}{2006}]{Shimasaku06}
{Shimasaku et al.} 2006, PASJ, 58, 313

\bibitem[\protect\citeauthoryear{{Sobral}, {Best}, {Geach}, {Smail} \& {et
  al.}}{{Sobral} et~al.}{2009a}]{Sobral}
{Sobral} D.,  {Best} P.~N.,  {Geach} J.~E.,  {Smail} I.,    {et al.} 2009a,
  MNRAS, 398, 75

\bibitem[\protect\citeauthoryear{{Sobral}, {Best}, {Geach}, {Smail} \& {et
  al.}}{{Sobral} et~al.}{2009b}]{Sobral2}
{Sobral} D.,  {Best} P.~N.,  {Geach} J.~E.,  {Smail} I.,    {et al.} 2009b,
  MNRAS, 398, L68

\bibitem[\protect\citeauthoryear{{Steidel}, {Giavalisco}, {Pettini},
  {Dickinson} \& {Adelberger}}{{Steidel} et~al.}{1996}]{steidel96}
{Steidel} C.~C.,  {Giavalisco} M.,  {Pettini} M.,  {Dickinson} M.,
  {Adelberger} K.~L.,  1996, ApJL, 462, L17+

\bibitem[\protect\citeauthoryear{{Tanvir et al.}}{{Tanvir et
  al.}}{2009}]{Tanvir}
{Tanvir et al.} 2009, Nature, 461, 1254

\bibitem[\protect\citeauthoryear{{Thommes} \& {Meisenheimer}}{{Thommes} \&
  {Meisenheimer}}{2005}]{Thommes05}
{Thommes} E.,  {Meisenheimer} K.,  2005, A\&A, 430, 877

\bibitem[\protect\citeauthoryear{{Willis} \& {Courbin}}{{Willis} \&
  {Courbin}}{2005}]{willis05}
{Willis} J.~P.,  {Courbin} F.,  2005, MNRAS, 357, 1348

\bibitem[\protect\citeauthoryear{{Willis}, {Courbin}, {Kneib} \&
  {Minniti}}{{Willis} et~al.}{2008}]{willis08}
{Willis} J.~P.,  {Courbin} F.,  {Kneib} J.-P.,    {Minniti} D.,  2008, MNRAS,
  384, 1039

\end{thebibliography}


\label{lastpage}

\end{document}